# Soft x-ray spectroscopy measurements of the *p*-like density of states of B in MgB$_2$ and evidence for surface boron oxides on exposed surfaces

T. A. Callcott,[1] L. Lin,[1] G. T. Woods,[1] G. P. Zhang,[1] J. R. Thompson,[1,2] M. Paranthaman,[2] and D. L. Ederer[3]
1. *The University of Tennessee, Knoxville, TN 37996*
2. *Oak Ridge National Laboratory, Oak ridge, TN 37831*
3. *Tulane University, New Orleans, LA 70118*

**Abstract**

Soft X-ray absorption and fluorescence measurements are reported for the *K*-edge of B in MgB$_2$. The measurements confirm a high density of B $p_{xy}(\sigma)$-states at the Fermi edge and extending to approximately 0.9 eV above the edge. A strong resonance is observed in elastic scattering through a core-exciton derived from out-of-plane $p_z(\pi^*)$-states. Another strong resonance, observed in both elastic and inelastic spectra, is identified as a product of surface boron oxides.

PACS numbers: 71.20.-b, 74.25.Jb, 78.70.DM, 78.70.En

Nagamatsu *et al.* have recently identified MgB$_2$ as a superconductor with a T$_c$ of 39 K.[1] This readily available compound consists of hexagonal, graphite-like layers of boron separated by layers of Mg. The Mg is partially ionized, donating electrons to the boron layer. The extra electrons help stabilize the $sp^2$ bonding of the boron layer, but do not completely fill the band associated with the in-plane $p_{xy}$-bonding orbitals. The resulting high density of $p_{xy}$-states at the Fermi level is believed to contribute to the high T$_c$ of the compound.

Soft x-ray absorption (SXA) and fluorescence (SXF) measurements at the B *K*-edge provide ideal tools for directly measuring the density of the crucial boron *p*-states in this compound. These spectroscopies involve dipole selected electronic transitions from and to a B 1*s* state, and thus provide a direct measure of the occupied and unoccupied *p*-like density of states of B in MgB$_2$. SXF is a filled state spectroscopy that maps the states below the Fermi edge and SXA an empty state spectroscopy mapping the states above the edge.[2] In this letter we provide measurements of these spectra and compare the results with the calculated *p*-like partial density of states (PDOS) of boron in MgB$_2$. The measurements confirm the accuracy of standard band structure codes for MgB$_2$ and the existence of a high PDOS derived from the B $p_{xy}$-orbitals at the Fermi edge.

Our measurements also indicate that surface oxides of boron are present in samples exposed to air so that careful protection of MgB$_2$ from ambient atmosphere may be required in applications. MgB$_2$ is relatively stable when exposed to air as is obvious from the very large volume of successful measurements that have been reported for powdered, sintered and compressed samples. Nevertheless, it is known to be hygroscopic, and reported work has indicated the formation of Mg oxides and hydroxides at the surface.[3] We observe resonance features in both the SXA and SXF spectra that indicate that surface boron oxides are also formed.

The sample was synthesized at Oak Ridge National Laboratory in powdered form, pressed and sintered to a density of about 70%, and characterized with magnetization measurements. The measured T$_c$ was 39 K.

In Figures 1a and 1b, we present the SXA absorption spectra of MgB$_2$ as measured by total fluorescent yield (TFY) and total electron yield (TEY) measurements, respectively. Three peaks are noted in the TFY spectra that are of special interest. Peak A at the threshold is derived from the high, unfilled $p_{xy}$-PDOS extending above the Fermi level. Peak B is associated with a strong resonance in the elastic scattering rather than from inelastic scattering processes that contribute to the SXF spectra. As discussed further at the end of this paper, we believe that peak C is associated with resonantly enhanced elastic and inelastic scattering from boron oxides.

The TEY spectrum in Figure 1b does not show a peak at the absorption threshold, but instead shows a shallow dip at the position of peak A. This is a result of the circumstances that the absorption threshold is superimposed on a strong background of electron emission from valence band electrons and the Mg *L* core levels, coupled with the competition of radiative and electronic de-excitation following the excitation of the B-1*s* core. Near threshold, SXF spectra are often dominated by a Resonant Inelastic X-ray Scattering (RIXS) process [2] in which coupling of incident and emitted photons creates strong resonant photon emission. When this becomes the strongly dominant de-excitation channel, few TEY electrons are produced by excitation of the 1*s* electrons at the *K* edge, and electronic excitation of valence and shallow core levels is reduced, generating a net decrease in electron emission. Clearly, in this case, the resonant radiative de-excitation channel is dominant and the TFY spectra is the appropriate spectrum to use for measuring structure in the density of states just above the threshold.

Emission spectra excited in the threshold region are shown in Figure 2a and in the region of the B and C resonances in Figure 2b. The sharp peaks with a width of about 0.4 eV are produced by elastic scattering of incident photons. Their widths provide a direct measure of the combined resolution of the beam-line monochromator, which delivers incident photons, and the emission spectrometer, which detects emitted photons. Their presence in the emission spectra provides a very accurate method of comparing the energy scales of the beam-line monochromator and emission spectrometer. The spectra located between about 180 and 188 eV is the normal emission spectra, which provides a measure of the filled state *p*-PDOS of B in $MgB_2$. The resonant enhancement of the upper shoulder in the spectrum excited at 187.75 eV is clear evidence for the presence of the RIXS process in the threshold region. All of the emission spectra excited at energies above about 190 eV, and excepting those excited at the resonance energy C, are essentially identical and provide a reliable measure of the *p*-PDOS of filled states at the boron site in $MgB_2$.

In Figure 3a, we plot the SXF and near threshold TFY absorption spectra together. The spectra overlap appropriately at the Fermi edge and provide a measure of the filled and empty *p*-like PDOS, modified primarily by instrumental broadening factors. We also plot (Figure 3b) a theoretical *p*-PDOS for B in $MgB_2$, calculated with the WIEN97 code.[4] The dotted line in the theoretical curve is obtained from the code and the solid curve has been obtained by broadening the curve with instrumental resolution functions appropriate for the beam-line monochromator and emission spectrometer. There is excellent general agreement between the experimental and theoretical curves. An analysis of the threshold peak in the absorption spectrum indicates that the $p_{xy}$-PDOS extends about 0.9 eV above the Fermi level.

In Figure 2b, the elastic peak is resonantly enhanced by a factor of 20 at 193 eV, the energy position of peak B of Figure 1a, while no change is observed in the emission spectrum. This results from elastic scattering through an intermediate core-exciton state derived from the localization of empty $p_z(\pi^*)$ orbitals. This resonant enhancement may be compared with the $\pi^*$ resonance in hexagonal BN, where the intensity enhancement in the elastic peak is an order of magnitude greater than observed here.[5] The difference may be attributed to the circumstance that the exciton lies within the band gap of *h*-BN and overlaps conduction band states in $MgB_2$.

In Figures 1a and 1b, we note that peak C is strongly enhanced in the TEY spectrum as compared with the TFY. Since TEY measurements are far more surface sensitive than TFY measurements, this result is consistent with our interpretation of this peak as being associated with surface boron oxides. In Figure 1c, we plot the absorption spectrum of $B_2O_3$, which shows a strong resonance very close to the position of peak C in the $MgB_2$ sample, further supporting this interpretation of the peak.

The spectrum excited at 194.8 eV, the energy of C in Figure 1, shows a moderate (X4) enhancement of the elastic peak, but also shows a low energy shoulder on the elastic peak and a broad enhancement of the SXF spectra centered at about 182 eV. As shown in Figure 4, both of these features are similar to features observed in the spectra of $B_2O_3$. Here, the $MgB_2$ spectrum excited at the C resonance is compared with the spectrum of $B_2O_3$ excited at its resonance peak. As a final test of the identification of peak C as being associated with boron oxide, we removed the sample from vacuum and left it in ambient air for twenty hours rather than in a desiccator. When returned to vacuum and re-measured, peak C had doubled in relative magnitude when compared to peaks A and B. Finally, we note that the contribution of oxides to the observed spectra were significant only when enhanced by strong resonance effects and did not contribute significantly to the reported spectra at other energies.

This research is supported by NSF grant DMR-9801804. Samples were prepared and characterized at ORNL supported by DOE contract DE-AC05-00OR-22725 Measurements were carried out at the Advanced Light Source at LBNL supported by DOE contract DE-A003-76SF00098.

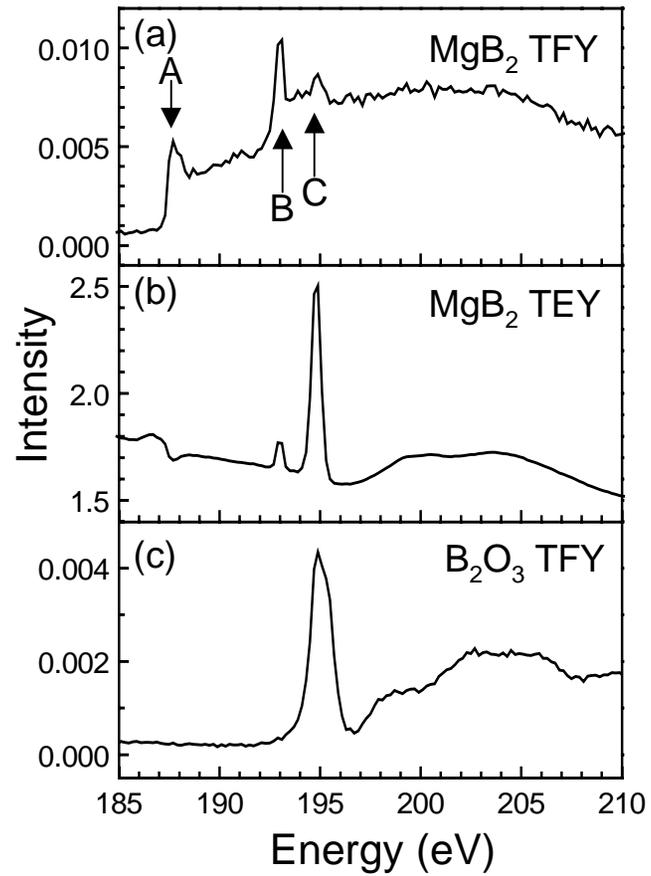

FIG. 1.(a) Total fluorescent yield of MgB$_2$ identifying three resonance features discussed in text. (b) Total electron yield of MgB$_2$. (c) Total electron yield of B$_2$O$_3$ showing resonance at peak coincident with peak C of curves (a) and (b).

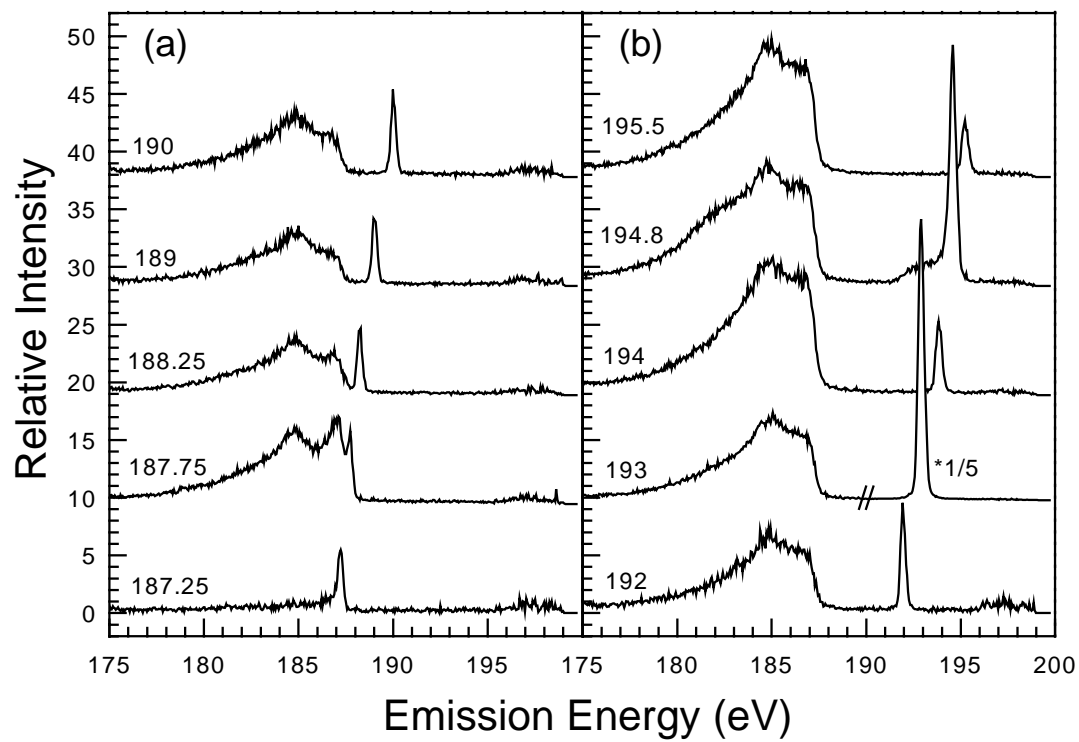

FIG. 2. SX fluorescence spectra excited near the boron *K* threshold. (b) SX fluorescence spectra excited near the core exciton resonances for MgB$_2$ and boron oxide.

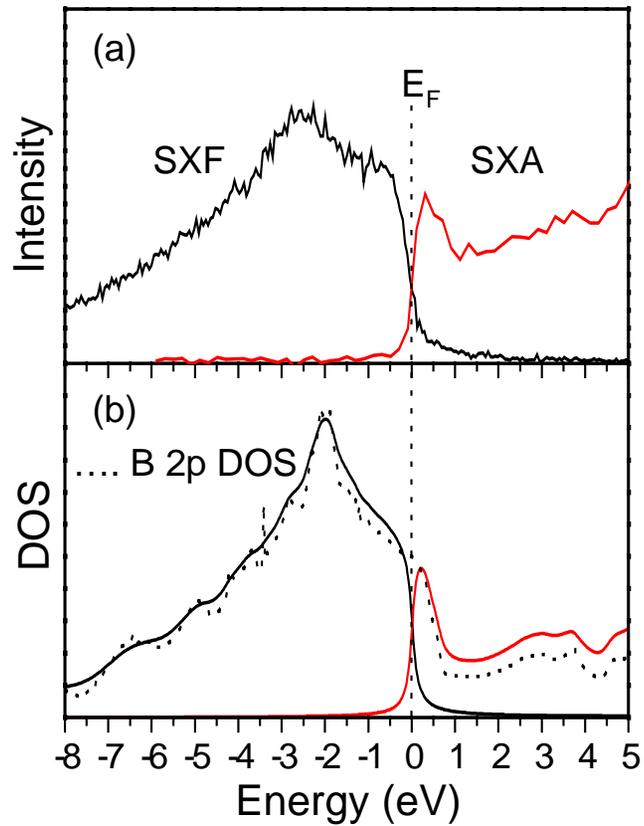

FIG. 3. Comparison of SX emission (SXF) and absorption (TFY) spectra with (b) calculated $p$ like partial density of states for boron in $MgB_2$. Solid curve is the theoretical curve (dotted) broadened by thermal, lifetime and instrumental functions.

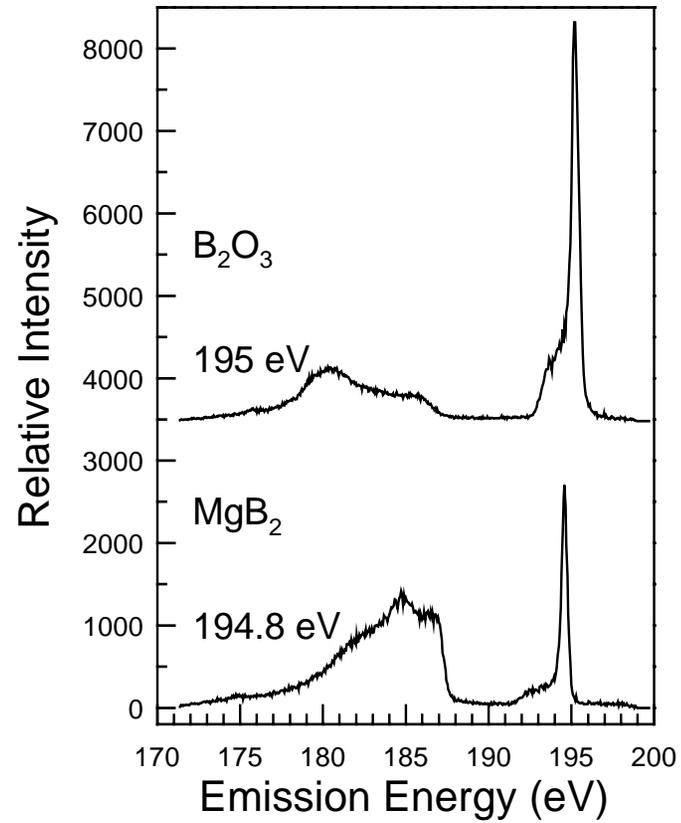

FIG. 4. Comparison of boron $K$ spectrum of $MgB_2$ excited at the "oxide" resonance (lower curve) with on-resonance excitation of $B_2O_3$ (upper curve).